\documentclass[twocolumn,showpacs,preprintnumbers,amsmath,amssymb,prb]{revtex4}
\usepackage{epsfig}
\usepackage{graphicx}
\usepackage{dcolumn}
\usepackage{bm}
\usepackage{color} 

\begin{document}

\title{
Origin of infrared peaks in the optical conductivity of ytterbium compounds
}
\author{Shin-ichi Kimura}
 \altaffiliation[Electronic address: ]{kimura@ims.ac.jp}
\affiliation{UVSOR Facility, Institute for Molecular Science, Okazaki 444-8585, Japan}
\affiliation{School of Physical Sciences, The Graduate University for Advanced Studies (SOKENDAI), Okazaki 444-8585, Japan}

\date{\today}
\begin{abstract} 
We have calculated optical conductivity [$\sigma(\omega)$] spectra of ytterbium compounds (YbAl$_3$, YbAl$_2$, YbCu$_2$Si$_2$, YbNi$_2$Ge$_2$, YbInCu$_4$, YbRh$_2$Si$_2$, YbIr$_2$Si$_2$, and YbB$_{12}$) based on the direct interband transition derived from first-principle band calculation and compared the results with the experimentally obtained $\sigma(\omega)$ spectra.
The spectral feature of a peak in the middle-infrared region (mid-IR peak) and a shoulder structure in the far-infrared region (far-IR shoulder) in the experimental $\sigma(\omega)$ spectra can be described by the band calculation with a common renormalization factor.
This result indicates that the infrared spectra of Yb compounds originate from the interband transition from the Yb~$4f$ state but that the Yb~$4f$ state shifts to the Fermi level with strong electron correlation.
\end{abstract}

%
\maketitle
%
%
In heavy fermion systems as well as mixed-valence materials, such as cerium (Ce), ytterbium (Yb), and uranium compounds, a characteristic spectral feature, in which free carriers absorption (and consequently Drude weight) due to quasiparticles with heavy carrier mass and long relaxation time, and middle-infrared peaks (mid-IR peaks), appear in optical conductivity [$\sigma(\omega)$] spectra in the far- and middle-infrared regions, respectively.~\cite{DG}
The origin of the heavy quasiparticle state is believed to be the hybridization effect between conduction and local $4f$ electrons, namely c-f hybridization, which can be explained by the periodic Anderson model (PAM).
In other words, the quasiparticle state due to c-f hybridization appears on the Fermi level ($E_{\rm F}$).~\cite{Hewson1993}
The c-f hybridization state as well as a heavy conduction band, which can be clearly explained by PAM, have been directly observed by a resonant angle-resolved photoemission spectroscopy of CeCoGe$_{1.2}$Si$_{0.8}$, a typical heavy fermion compound.~\cite{Im2008}

On the other hand, although mid-IR peaks clearly appear in the $\sigma(\omega)$ spectra of many Ce and Yb compounds,~\cite{Degiorgi1999} their origin has remained a long-debated issue.
One explanation of the origin of the mid-IR peaks is the optical transition between the bonding and antibonding states of the c-f hybridization gap described by PAM.~\cite{Dordevic2001,Hancock2004,Hancock2006,Okamura2007}
In this case, the energy of the mid-IR peak should correspond to the hybridization intensity and be proportional to the square root of the Kondo temperature ($T_{\rm K}$).~\cite{Degiorgi2001}
Another explanation based on band structure has been proposed; i.e., that mid-IR peaks originate from a characteristic band structure in which the $4f$ state is located near $E_{\rm F}$.~\cite{Saso2004,Kuroiwa2007,Kimura2009}
In particular, both the energy position and spectral shape of the mid-IR peaks of Ce$X_3$ ($X$~=~Pd, Sn, In) compounds can be explained by first-principle band structure calculation with the spin-orbit splitting (SOS).~\cite{Kimura2009}
In Yb compounds, especially YbAl$_3$, however, the experimental $\sigma(\omega)$ spectra can be explained by band calculation in which the $4f$ state is slightly shifted toward $E_{\rm F}$.~\cite{Kuroiwa2007}
The origin of the energy shift of the Yb~$4f$ state is not clear, even though the energy of the Ce~$4f$ state in Ce$X_3$ is fixed for comparison.~\cite{Kimura2009}

In this paper, we describe our investigation into the origin of the mid-IR peak in Yb compounds and also the origin of the low-energy shift of the Yb~$4f$ state if the origin is the band structure.
We calculated $\sigma(\omega)$ spectra of heavy fermion Yb compounds (YbAl$_3$, YbAl$_2$, YbCu$_2$Si$_2$, YbNi$_2$Ge$_2$, YbInCu$_4$, YbRh$_2$Si$_2$, and YbIr$_2$Si$_2$) and a Kondo insulator (YbB$_{12}$) from first-principle band calculation with SOS and compared the results with the experimental $\sigma(\omega)$ spectra.
It was found that all of the calculated $\sigma(\omega)$ spectra reproduce the experimental spectra as the optical transition from the Yb~$4f_{7/2}$ state to the unoccupied state mainly originating from the Yb~$5d$ state when the photon energy of all of the calculated spectra is reduced.
This result indicates that the mid-IR peaks in Yb compounds also originate from the characteristic band structure.
The energy shift can be regarded as a common renormalization factor due to the strong electron correlation of the Yb~$4f$ state.

%
\begin{table*}[t]
\caption{Space group, lattice constant and Kondo temperature of calculated Yb compounds.}
\label{LatticeParameter}
\begin{tabular}{@{\hspace{\tabcolsep}\extracolsep{\fill}}cccc} \hline
		& Space group		& Lattice constant (pm)				& Kondo temperature (K)	\\ \hline
YbAl$_3$	& No.~221, $Pm-3m$	& $a,b,c=420.0$~\cite{Kuroiwa2007}		& 670~\cite{Ebihara2002}	\\
YbAl$_2$	& No.~227, $Fd-3m$	& $a,b,c=788.5$~\cite{Lee2000}			& 2000~\cite{Dallera2003}	\\
YbCu$_2$Si$_2$	& No.~139, $I4/mmm$	& $a,b=391.2$, $c=997.0$~\cite{Winkelmann1999}	& 40~\cite{Joyce1996}		\\
YbNi$_2$Ge$_2$	& No.~139, $I4/mmm$	& $a,b=400.1$, $c=973.3$~\cite{Knebel2001}	& 30~\cite{Budko1999}	\\
YbInCu$_4$	& No.~216, $F-43m$	& $a,b,c=712.3$~\cite{Takegahara1990}		& 400~\cite{Garner2000}	\\
YbRh$_2$Si$_2$	& No.~139, $I4/mmm$	& $a,b=400.7$, $c=985.8$~\cite{Trovarelli2000}	& 25~\cite{Joerg2003}	\\
YbIr$_2$Si$_2$	& No.~139, $I4/mmm$	& $a,b=403.2$, $c=982.6$~\cite{Hossain2005}	& 40~\cite{Hossain2005}	\\
YbB$_{12}$	& No.~225, $Fm-3m$	& $a,b,c=746.4$~\cite{Yanase1992}		& 240~\cite{Iga1999}	\\ \hline
\end{tabular}
\end{table*}
LDA band structure calculation was performed by the full potential linearized augmented plane wave plus local orbital (LAPW+lo) method including SOS implemented in the {\sc Wien2k} code.~\cite{WIEN2k}
The lattice parameters listed in Table~\ref{LatticeParameter} reported in the literature for YbAl$_3$,~\cite{Kuroiwa2007} YbAl$_2$,~\cite{Lee2000} YbCu$_2$Si$_2$,~\cite{Winkelmann1999} YbNi$_2$Ge$_2$,~\cite{Knebel2001} YbInCu$_4$,~\cite{Takegahara1990} YbRh$_2$Si$_2$,~\cite{Trovarelli2000} $I$-type YbIr$_2$Si$_2$~\cite{Hossain2005} and YbB$_{12}$~\cite{Yanase1992} were used for the calculation.
$R_{MT}K_{max}$ (the smallest MT radius multiplied by the maximum $k$ value in the expansion of plane waves in the basis set), which determines the accuracy of the basis set used, was set at 7.0.
The total number of Brillouin zones was sampled with 40,000~$k$-points.
The obtained band structures of YbAl$_3$, YbAl$_2$, YbInCu$_4$, YbRh$_2$Si$_2$, and YbB$_{12}$ are consistent with the previous band calculations with SOS reported by other groups.~\cite{Kuroiwa2007,Lee2000,Antonov2000,Yanase1992}
The $T_{\rm K}$ is also listed in Table~\ref{LatticeParameter} for reference.
Note that the LDA band calculation of YbB$_{12}$ produces a semi-metallic band structure, which is not consistent with the experimental semiconducting character.
However, the purpose of this letter is to investigate the origin of mid-IR peaks in Yb compounds from the first-principle band calculation, therefore we used the semi-metallic band structure from the LDA calculation.

The $\sigma(\omega)$ spectra were derived from a function as follows, in which direct interband transitions were assumed;~\cite{Ant04}
\begin{equation}
\label{eq:1}
\hat{\sigma}(\omega) = \frac{\pi e^2}{m_0^2 \omega} \sum_{\vec{k}} \sum_{n n'} \frac{|\langle n' \vec{k}|\vec{e} \cdot \vec{p}|n \vec{k}\rangle |^{2}}{\omega - \omega_{n n'}(\vec{k})+i\Gamma} \times \frac{f(\varepsilon_{n\vec{k}})-f(\varepsilon_{n'\vec{k}})}{\omega_{n n'}(\vec{k})}.
\end{equation}
Here, the $|n' \vec{k}\rangle$ and $|n \vec{k}\rangle$ states denote the unoccupied and occupied states, respectively, $\vec{e}$ and $\vec{p}$ are the polarization of light and the momentum of the electron, respectively, $f(\varepsilon_{n\vec{k}})$ is the Fermi-Dirac distribution function, $\hbar\omega_{n n'}=\varepsilon_{n\vec{k}}-\varepsilon_{n'\vec{k}}$ is the energy difference between the unoccupied and occupied states, and $\Gamma$ is the lifetime.
In the calculation, $\Gamma$~=~1~meV was assumed.
The experimental $\sigma(\omega)$ spectra below the temperature of 10~K already reported in the literature for YbAl$_3$,~\cite{Okamura2004} YbAl$_2$,~\cite{Okamura2007} YbCu$_2$Si$_2$,~\cite{Okamura2007} YbNi$_2$Ge$_2$,~\cite{Okamura2007} YbInCu$_4$,~\cite{Okamura2007-2} YbRh$_2$Si$_2$,~\cite{Kimura2006} $I$-type YbIr$_2$Si$_2$,~\cite{Sichelschmidt2008} and YbB$_{12}$~\cite{Okamura2005} were used for comparison with the calculated spectra.
Since the experimental $\sigma(\omega)$ spectra of single-crystalline YbCu$_2$Si$_2$, YbNi$_2$Ge$_2$, YbRh$_2$Si$_2$, and YbIr$_2$Si$_2$ have been measured in the $ab$-plane, $\sigma(\omega)$ calculation along the $a$-axis was performed.

%
\begin{figure}[t]
\begin{center}
\includegraphics[width=0.35\textwidth]{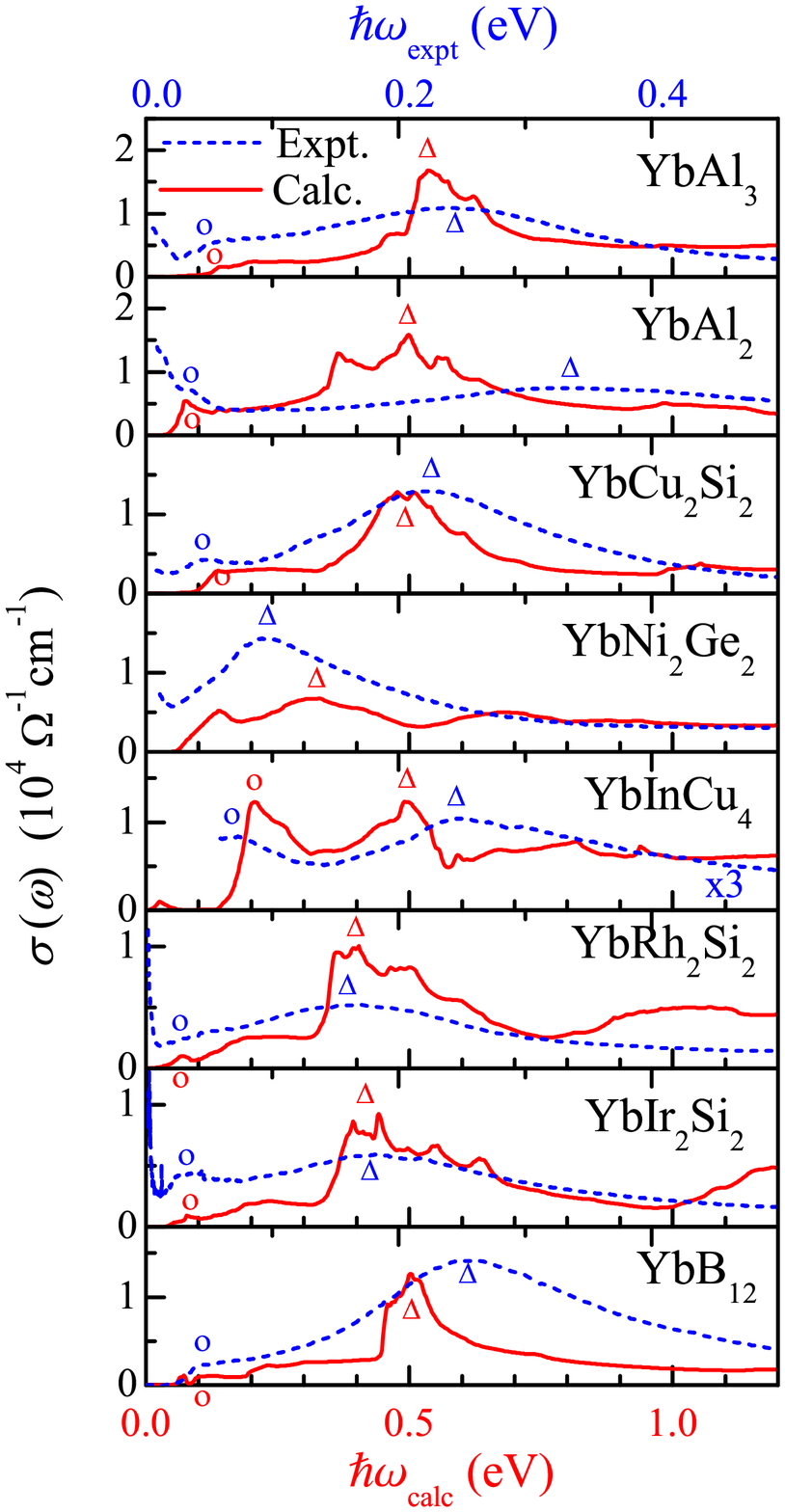}
\end{center}
\caption{(Color online)
Calculated optical conductivity [$\sigma(\omega)$] spectra compared with experimental spectra of YbAl$_3$,~\cite{Okamura2004} YbAl$_2$,~\cite{Okamura2007} YbCu$_2$Si$_2$,~\cite{Okamura2007} YbNi$_2$Ge$_2$,~\cite{Okamura2007} YbInCu$_4$,~\cite{Okamura2007-2} YbRh$_2$Si$_2$,~\cite{Kimura2006} YbIr$_2$Si$_2$,~\cite{Sichelschmidt2008} and YbB$_{12}$.~\cite{Okamura2005}
The $\sigma(\omega)$ spectra of YbCu$_2$Si$_2$, YbNi$_2$Ge$_2$, YbRh$_2$Si$_2$, and YbIr$_2$Si$_2$ were calculated along the $a$-axis, in the same way as the experimental spectra.
The horizontal scales of the experimental and calculated spectra are located at the top and bottom, respectively.
Note that the horizontal scale of the calculated $\sigma(\omega)$ spectra is reduced by a factor of 0.42 (=~0.5/1.2) compared with that of the experimental spectra.
The vertical scales of the calculated and experimental $\sigma(\omega)$ spectra are identical except in the case of YbInCu$_4$, in which the experimental spectrum is three times as large as the calculated spectrum.
The mid-IR peaks and far-IR shoulders of the experimental and calculated spectra are marked by open triangles and open circles, respectively.
}
\label{OC}
\end{figure}
\begin{figure}[t]
\begin{center}
\includegraphics[width=0.35\textwidth]{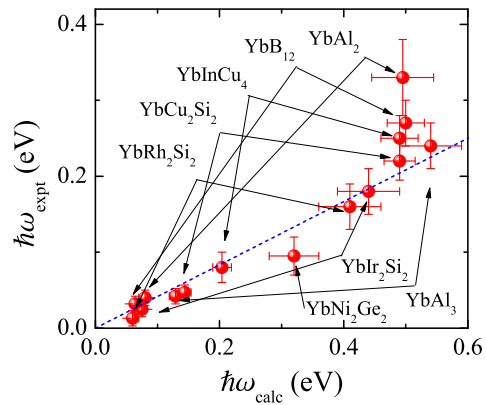}
\end{center}
\caption{(Color online)
Experimental energies ($\hbar\omega_{expt}$) of the mid-IR peaks and far-IR shoulders as a function of the corresponding calculated energies ($\hbar\omega_{calc}$) of Yb compounds (solid circles).
The error bars show the width of the peak top in both the experimental and calculated spectra.
The dashed line indicates $\hbar\omega_{expt}=0.42 \times \hbar\omega_{calc}$, which is the reducing ratio of the photon energy in Fig.~\ref{OC}
}
\label{PeakPlot}
\end{figure}
The calculated $\sigma(\omega)$ spectra of the Yb compounds compared with previously reported experimental spectra are shown in Fig.~\ref{OC}.
The experimental spectra have the peculiar features of the mid-IR peak (open triangles), as already pointed out, of the shoulder structure in the far-infrared region below 0.1~eV (far-IR shoulder, open circles), and of the Drude weight of quasiparticles below 0.02~eV.
Since the calculation was only performed for the interband transition, the Drude weight of quasiparticles is not discussed here.
The calculated $\sigma(\omega)$ spectra mainly originate from the interband transition from the Yb~$4f_{7/2}$ to the unoccupied state including the Yb~$5d$ state.
The experimental and calculated horizontal scales are set at 0--0.5~eV (top axis) and 0--1.2~eV (bottom axis), respectively.
From this comparison, the experimental $\sigma(\omega)$ spectra can be seen to be roughly explained by the calculation;
for example, in YbAl$_3$, the experimental mid-IR peak and far-IR shoulder at around 0.25~eV and 0.05~eV correspond to those at around 0.6~eV and 0.1~eV in the calculation, respectively.
In other Yb compounds, not only the mid-IR peak but also the far-IR shoulder in the experimental spectra are explained well by the calculated $\sigma(\omega)$ spectra.
The experimental peak height is also reproduced by the band calculation except in the case of YbInCu$_4$.~\cite{YbInCu4}
For example, the heights of the mid-IR peaks of YbAl$_3$, YbCu$_2$Si$_2$, and YbB$_{12}$ exceed $10^4 \Omega^{-1}{\rm cm}^{-1}$, which corresponds to the calculated height, while, those of YbRh$_2$Si$_2$ and YbIr$_2$Si$_2$ are about $5\times10^3 \Omega^{-1}{\rm cm}^{-1}$, which can be reproduced by the calculation.
The calculated $\sigma(\omega)$ spectra can therefore successfully reproduce the experimental spectra.
In other words, the infrared spectra of Yb compounds, especially the characteristic mid-IR peaks and far-IR shoulders, originate from the band structure.
However, the horizontal scale of the calculated spectra is reduced by 0.42 (=~0.5/1.2).
The meaning of the reducing ratio is discussed next.

The energies of the mid-IR peaks and far-IR shoulders derived from the experimental $\sigma(\omega)$ spectra shown in Fig.~\ref{OC} are plotted as a function of the calculated photon energy of the peaks and shoulders in Fig.~\ref{PeakPlot}.
The dashed line in the figure is the reducing ratio ($\hbar\omega_{expt} = 0.42 \times \hbar\omega_{calc}$) of the photon energy in the calculated $\sigma(\omega)$ spectra shown in Fig.~\ref{OC}.
This figure indicates that almost all of the mid-IR peaks and far-IR shoulders except for the mid-IR peak of YbAl$_2$ are located on the dashed line.
This means that the experimental $\sigma(\omega)$ spectra are similarly renormalized from the spectra derived from the band structure calculation in proportion to the photon energy.
Since both the mid-IR peaks and far-IR shoulders originate from the interband transition from the Yb~$4f_{7/2}$ state, the energy shift can be explained by the renormalization of the Yb~$4f_{7/2}$ state due to the strong electron correlation.
The Yb~$4f_{7/2}$ peak in the LDA calculation of YbAl$_3$ is actually located at about 200~meV from the $E_{\rm F}$~\cite{Kuroiwa2007}, but it has been observed at 43~meV in a photoemission spectrum.~\cite{Suga2005}
In YbIr$_2$Si$_2$ and YbRh$_2$Si$_2$, the Yb~$4f_{7/2}$ dispersion is also pushed up to $E_{\rm F}$ from that of the LDA calculation.~\cite{Dan2006,Dan2007,Wigger2007}
Therefore, since the Yb~$4f_{7/2}$ peak is indeed strongly renormalized, the same renormalization effect appears in the $\sigma(\omega)$ spectra.

The renormalization effect of the Yb~$4f_{7/2}$ peak can be described by the renormalized perturbation theory based on PAM.~\cite{Hewson1993}
Due to this theory, the one-particle Green's function for the $4f$ electron is written as
\begin{equation}
\label{eq:2}
G_f(\varepsilon)=\frac{1}{\varepsilon-\varepsilon_f+i\Delta(\varepsilon)-\Sigma(\varepsilon)}=\frac{z}{\varepsilon-\tilde{\varepsilon}_f+i\tilde{\Delta}(\varepsilon)},
\end{equation}
where $\varepsilon_f$ and $\Delta(\varepsilon)$ are the energy and width of the non-interacted $4f$ level, respectively, and $\Sigma(\varepsilon)$ the self-energy of the $4f$ state.
$z$, the wave function renormalization factor, is give by
\begin{equation}
\label{eq:3}
z=\left(\left.1-\frac{d\Sigma(\varepsilon)}{d\varepsilon}\right|_{\varepsilon=0}\right)^{-1},
\end{equation}
and $\tilde{\varepsilon}_f=z\varepsilon_f$ and $\tilde{\Delta}(\varepsilon)=z\Delta(\varepsilon)$.

Since the spectral function of the Yb~$4f_{7/2}$ peak [$\rho_f(\varepsilon)$] is derive from $\rho_f(\varepsilon)=-\frac{1}{\pi}{\rm Im}G_f(\varepsilon)$,
the renormalized Yb~$4f_{7/2}$ peak ($\tilde{\varepsilon}_f$) shifts to the low binding energy side from the non-interacted $4f$ state ($\varepsilon_f$) by $z$.
If the mid-IR peaks originate from the interband transition from the Yb~$4f_{7/2}$ to unoccupied states, $\varepsilon_f$ and $\tilde{\varepsilon}_f$ can be regarded to be $\hbar\omega_{calc}$ and $\hbar\omega_{expt}$, respectively.
Therefore $\hbar\omega_{expt}$ must be equal to $z\hbar\omega_{calc}$, which is consistent with the linear relation in Fig.~\ref{PeakPlot}.
Then the renormalization factor $z$ becomes 0.42 from the dashed line in Fig.~\ref{PeakPlot}, the self-energy $\Sigma(\varepsilon)=-1.4\varepsilon$ derived from Eq.~\ref{eq:3}, and the effective mass ($m^*=m_0/z$) becomes $2.4m_0$, where $m_0$ is the free electron mass.
The renormalization effect is consistent with the photoemission spectrum of YbB$_{12}$, because the Kondo peak originating from the Yb~$4f_{7/2}$ state shifts close to $E_{\rm F}$ from the original Yb~$4f_{7/2}$ state derived from the band structure calculation due to the renormalization effect.~\cite{Susaki1997}
The same effect is considered to appear in the $\sigma(\omega)$ spectra of the other Yb compounds speculated from Fig.~\ref{PeakPlot}.
This result is consistent with the fact that the real part of self-energy is proportional to the energy near $E_{\rm F}$~\cite{Vries1994} and actually within a few tenths of an eV.~\cite{Susaki1997}

So far, we have clearly shown that not only the mid-IR peaks but also the far-IR shoulders of Yb compounds originate from the band structure as well as the interband transition from the Yb~$4f_{7/2}$ to the unoccupied state.
Our previous study on Ce$X_3$ ($X$~=~Pd, Sn, In) also pointed out that the mid-IR peaks can be explained by the band structure calculation.~\cite{Kimura2009}
Therefore, it is concluded that \textit{the origin of the infrared $\sigma(\omega)$ peaks of both Ce and Yb compounds is the band structure} in which the local $4f$ state is located very near $E_{\rm F}$.
The Yb compounds need the renormalization, however, despite the fact that no correction is needed in the Ce compounds.
The reason for this is not clear, but the difference in the $4f$ occupation number may play an important role.

Next, the reason why the experimental mid-IR peak of YbAl$_2$ is located at the higher energy side than the dashed line in Fig.~\ref{PeakPlot} is discussed.
As shown in Table~\ref{LatticeParameter}, the $T_{\rm K}$ of YbAl$_2$ is much higher than those of the other compounds.
Both the high $T_{\rm K}$ and the intermediate valence~\cite{Dallera2003} indicate the weak electron correlation of the Yb~$4f$ state in YbAl$_2$.
Hence, the renormalization factor is speculated to be close to 1.
Actually, $z$ of YbAl$_2$ can be evaluated to be about 0.73 from Fig.~\ref{PeakPlot}, which is larger than $z$ in other compounds.
Therefore $z$ is concluded to increase with $T_{\rm K}$, which is consistent with the renormalized perturbation theory.~\cite{Hewson1993}

The empirical common renormalization discussed in this work hides the physical meaning of the correlation effects that are generally associated with the conventional c-f hybridization and captured by several theoretical treatments like the slave boson or the dynamical mean field approach.
To clarify the basis of the renormalization, further theoretical approaches are needed.~\cite{Rozenberg1996,Vid2003,Franco2009}

%
To summarize, we derived the $\sigma(\omega)$ spectra of Yb compounds (YbAl$_3$, YbAl$_2$, YbCu$_2$Si$_2$, YbNi$_2$Ge$_2$, YbInCu$_4$, YbRh$_2$Si$_2$, YbIr$_2$Si$_2$, and YbB$_{12}$) in the infrared region from first-principle band calculation 
 and compared them with the experimental spectra.
The calculated $\sigma(\omega)$ spectra are consistent with the experimental spectra after the common renormalization even though the calculated Yb compounds have different crystal structures and different Kondo temperatures.
This result indicates that the infrared spectra including the mid-IR peaks and far-IR shoulders originate from the interband transition from the strongly correlated Yb~$4f_{7/2}$ state to the unoccupied state.
The mid-IR peaks and far-IR shoulders reflect the renormalization effect of the Yb~$4f_{7/2}$ states due to the strong electron correlation.
Although strong electronic correlations cannot be treated properly by LDA calculation, they seem to be reasonable for Yb-based heavy fermion systems results which fit the data when the common renormalization is applied.

%
We would like to thank Prof. Okamura and Dr. Sichelschmidt for their fruitful discussion and comments.
This work was partly supported by a Grant-in-Aid for Scientific Research (B) (No.~18340110) from the MEXT of Japan.

%

\begin{thebibliography}{99}
\bibitem{YbInCu4} The intensity of the experimental $\sigma(\omega)$ spectrum of YbInCu$_4$ is much smaller than those of the other Yb compounds. The reason might be the different surface preparation method, because the intensity of the $\sigma(\omega)$ spectrum with the cleaved surface used here is lower than that with the polished surface.~\cite{Okamura2007-2} The intensity of the $\sigma(\omega)$ spectrum measured by another group~\cite{Garner2000} is also higher.
\end{thebibliography}
\end{document}